\documentclass[traditabstract]{aa}

\newcommand{\Msun}{\ensuremath{\,{\rm M}_\odot}}                  
\newcommand{\Rsun}{\ensuremath{\,{\rm R}_\odot}}                  
\newcommand{\Lsun}{\ensuremath{\,{\rm L}_\odot}}                  
\newcommand{\Teff}{\ensuremath{T_{\rm eff}}}                      
\newcommand{\Porb}{\ensuremath{P_{\rm orb}}}                      
\newcommand{\Vsini}{\ensuremath{V \sin i}}                        
\newcommand{\Mbol}{\ensuremath{M_{\rm bol}}}                      
\newcommand{\kms}{\,km\,s$^{-1}$}                                 
\newcommand{\Vsync}{\ensuremath{V_{\rm synch}}}                   
\newcommand{\chir}{\ensuremath{\chi_\nu^{\,2}}}                   
\newcommand{\mc}[1]{\multicolumn{2}{c}{#1}}
\newcommand{\reff}[1]{{#1}}
\newcommand{\refff}[1]{{#1}}
\newcommand{\er}[3]{\ensuremath{#1^{+#2}_{-#3}}}

\usepackage{graphicx}
\usepackage{txfonts}
\usepackage{natbib}
\bibpunct{(}{)}{;}{a}{}{,}

\begin{document}

\title{The solar-type eclipsing binary system LL\,Aquarii}

\author{J.\ Southworth}

\institute{Astrophysics Group, Keele University, Staffordshire, ST5 5BG, UK \\ \email{astro.js@keele.ac.uk}}

\date{Received ????; accepted ????}

\abstract{The eclipsing binary LL\,Aqr consists of two late-type stars in an eccentric orbit with a period of 20.17\,d. We use an extensive light curve from the SuperWASP survey augmented by published radial velocities and $UBV$ light curves to measure the physical properties of the system. The primary star has a mass of $1.167 \pm 0.009$\Msun\ and a radius of $1.305 \pm 0.007$\Rsun. The secondary star is an analogue of the Sun, with a mass and radius of $1.014 \pm 0.006$\Msun\ and $0.990 \pm 0.008$\Rsun\ respectively. The system shows no signs of stellar activity: the upper limit on spot-induced rotational modulation is 3\,mmag, it is slowly rotating, has not been detected at X-ray wavelengths, and the calcium H and K lines exhibit no emission. Theoretical stellar models provide a good match to its properties for a sub-solar metal abundance of $Z = 0.008$ and an age of 2.5\,Gyr. Most low-mass eclipsing binary systems are found to have radii larger than expected from theoretical predictions, blamed on tidally-enhanced magnetic fields in these short-period systems. The properties of LL\,Aqr support this scenario: it exhibits negligible tidal effects, shows no signs of magnetic activity, and matches theoretical models well.}


\maketitle

\section{Introduction}                                                                                                              \label{sec:intro}

The eclipsing nature of LL\,Aqr was detected by using the {\it Hipparcos} satellite \citep{Perryman+97aa,Kazarovets+99ibvs}, photometry from which showed two eclipses of clearly different depth separated by roughly two years. The spectral type of the system has been classified as G1\,V \citep{HoukSwift99book}.

\citet{OteroDubovsky04ibvs} used the {\it Hipparcos} observations plus photometry from the NSVS \citep{Wozniak+04aj} and ASAS \citep{Pojmanski97aca} surveys to determine an orbital period of 20.1784\,d. The orbit is quite eccentric, resulting in a secondary eclipse near phase 0.3. The eclipses are short compared to the orbital period, so LL\,Aqr is a detached eclipsing binary (dEB).

\citet[][hereafter I08]{Ibanoglu+08mn2} presented extensive $UBV$ photometry which nevertheless does not cover the ingress or egress of secondary minimum. They also obtained twelve spectra from which radial velocities (RVs) were measured -- ten from the 91\,cm telescope at the Catania Astrophysical Observatory, Italy, and two from the 150\,cm telescope at T\"UBITAK National Observatory, Turkey. I08 used these data to determine the physical properties of the system, obtaining masses and radii to precisions of 5\% and 1.5\%, respectively. The primary is a slightly evolved 1.2\Msun\ star whereas the secondary is a solar twin. A third component was also detected in the two T\"UBITAK spectra, with an RV close to the systemic velocity of the dEB, but not in the ten Catania spectra.

Most recently, \citet[][hereafter G13]{Griffin13obs} presented 25 high-precision RVs for the two eclipsing components obtained with the Cambridge CORAVEL instrument, which directly observes cross-correlation functions of objects using a spectrum of Arcturus as the template \citep{Griffin67apj}. No trace of the putative third component was identified. The RVs of the two eclipsing stars are of sufficient quality to give their masses to better than 1\%, making a definitive characterisation of the two stars possible if combined with sufficient photometry. Numerous photometric observations are available from the SuperWASP database.

Analyses of low-mass dEBs usually yield radii which are too large to match the predictions of theoretical stellar models \citep{Hoxie73aa,Lopez07apj}, a phenomenon which is attributed to enhanced stellar activity due to tidal effects. The wide separation of the two components of LL\,Aqr means they should be relatively unaffected by tides, so their properties are important indicators of the reliability of theoretical models in this mass range. In this work we present a determination of the physical properties of LL\,Aqr.


\section{Observations}                                                                                                               \label{sec:obs}

\begin{figure*} \includegraphics[width=\textwidth,angle=0]{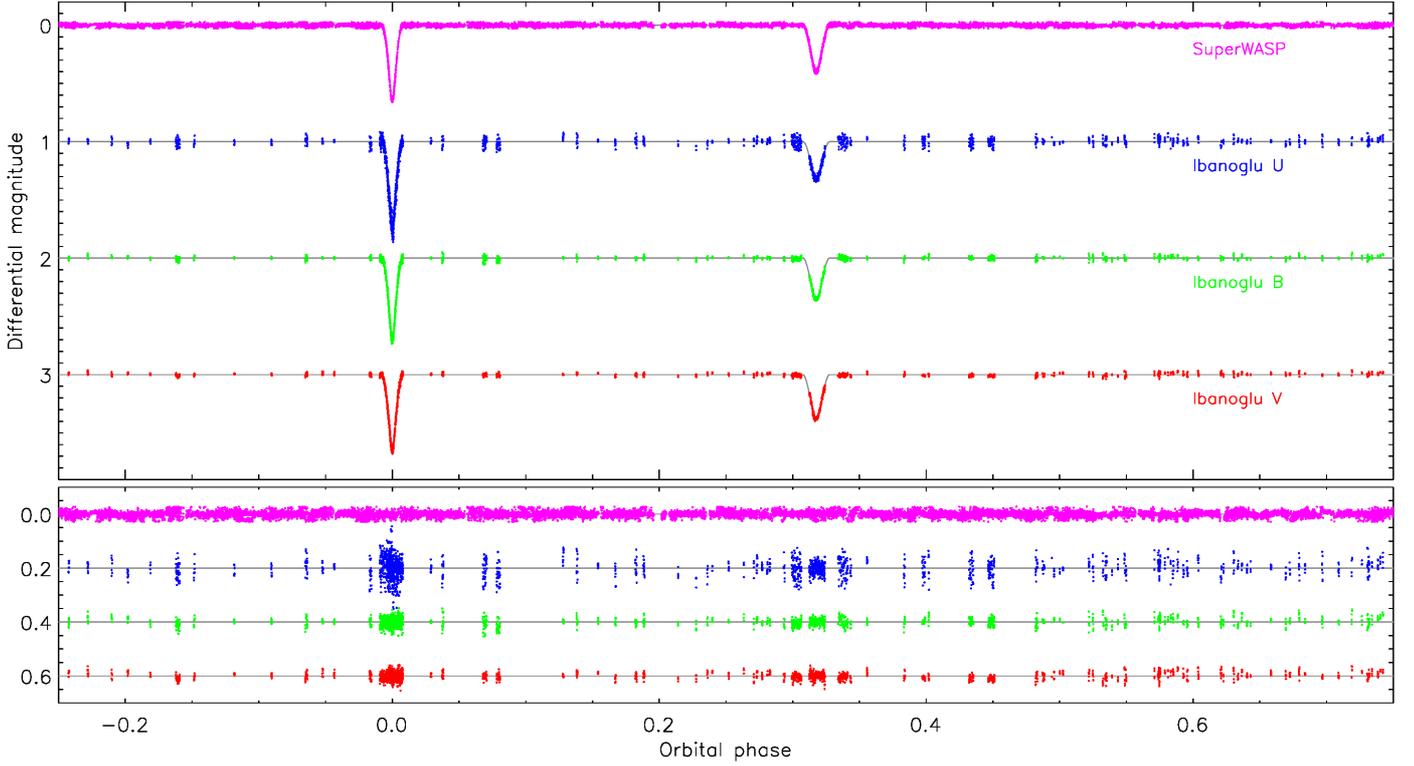}
\caption{\label{fig:lcall} Full light curves of LL\,Aqr from SuperWASP and 
I08. \reff{The {\sc jktebop} best fits are plotted as black lines and the 
residuals of the fits are shown in the lower panel.}} \end{figure*}

An extensive light curve of LL\,Aqr has been obtained by the SuperWASP consortium \citep{Pollacco+06pasp} in the course of a survey to detect new transiting extrasolar planets. Over 25\,000 observations have been obtained to date, yielding coverage of all orbital phases and the potential for a precise measurement of the physical properties of the system. The observations were taken through the SuperWASP wide filter, which has a response function close to that of Gunn $g$+$r$. Data were obtained using four of the cameras: 14\,576 points from camera 141, 87 from camera 142, 1054 from camera 147 (all part of the SuperWASP-North installation at La Palma, Spain), and finally 9364 points from camera 223 (at the SuperWASP-South installation, Sutherland, South Africa). Cameras 142 and 147 did not obtain any data during eclipse, so we did not involve these data in the analysis.

After inspecting the raw and detrended light curves, we selected those which had been detrended using the {\sc sysrem} algorithm \citep{Tamuz++05mn} for further study. The 23\,940 datapoints from cameras 141 and 223 have a median photometric precision of approximately 0.01\,mag, but have many outliers due to weather or technical issues. We therefore calculated a preliminary fit to the light curve and iteratively rejected those datapoints lying greater than 3$\sigma$ from the best fit, finishing with 21\,362 datapoints. \reff{This was done using the {\sc jktebop} code (see below) and ten iterations.} We tested the use of a 4$\sigma$ clipping threshold, and found that this did not have a significant effect on our results but caused the retention of several groups of datapoints taken on the same nights and systematically offset in magnitude from the remaining data. \reff{The full photometric data are given in Table\,\ref{tab:lcdata}.}

Finally, we obtained the published $UBV$ light curves from I08 and RVs from \citet{Griffin13obs} for inclusion in our analysis. \reff{We did not include the RVs measured by I08 due to their lower number and precision compared to those from \citet{Griffin13obs}.} We also searched the literature for measured times of minimum light in order to refine the orbital period of LL\,Aqr, finding \reff{six. Table\,\ref{tab:minima} gives these times of minimum along with their residual versus the best fit obtained below.} The full light curves from SuperWASP and I08 are shown in Fig.\,\ref{fig:lcall}.

\begin{table} \centering \caption{\label{tab:lcdata} 
\reff{SuperWASP photometric data for LL\,Aqr. The 
data are available in their entirety at the CDS.}}
\begin{tabular}{l c c c c} \hline \hline
Camera & HJD(UTC) & Magnitude & Error \\
\hline
141 & 2454733.411620 & $-$5.7100 & 0.0055 \\
141 & 2454733.415475 & $-$5.7163 & 0.0051 \\
141 & 2454733.415914 & $-$5.7189 & 0.0045 \\
141 & 2454733.419769 & $-$5.7193 & 0.0024 \\
141 & 2454733.420231 & $-$5.7230 & 0.0024 \\
\hline \end{tabular} \end{table}

\begin{table} \centering \caption{\label{tab:minima} 
\reff{Published times of minimum light for LL\,Aqr.}}
\begin{tabular}{l@{\,$\pm$\,}l r r r} \hline \hline
\mc{Time of minimum (HJD)} & Cycle & $O-C$ (d) & Reference \\
\hline
2448762.552  & 0.005 & $-$314.0 & $-$0.0045 & 1 \\   
2453968.5644 & 0.002 &  $-$56.0 &    0.0008 & 2 \\   
2454049.2766 & 0.001 &  $-$52.0 & $-$0.0003 & 2 \\   
2454358.3617 & 0.001 &  $-$36.5 &    0.0000 & 2 \\   
2454392.3095 & 0.001 &  $-$35.0 &    0.0011 & 2 \\   
2454735.334  & 0.005 &  $-$18.0 & $-$0.0058 & 3 \\   
\hline \end{tabular}  
\tablebib{(1) \citet{OteroDubovsky04ibvs}; (2) I08; (3) \citet{Bakan09bavsr}.}
\tablefoot{An integer cycle number refers to a time of primary eclipse 
whereas a half-integer cycle number is for a secondary eclipse. \refff{$O-C$ 
gives the residual versus the best fit obtained in Section.\,\ref{sec:anal}.}}
\end{table}


\section{Data analysis}                                                                                                              \label{sec:anal}

\begin{figure} \includegraphics[width=\columnwidth,angle=0]{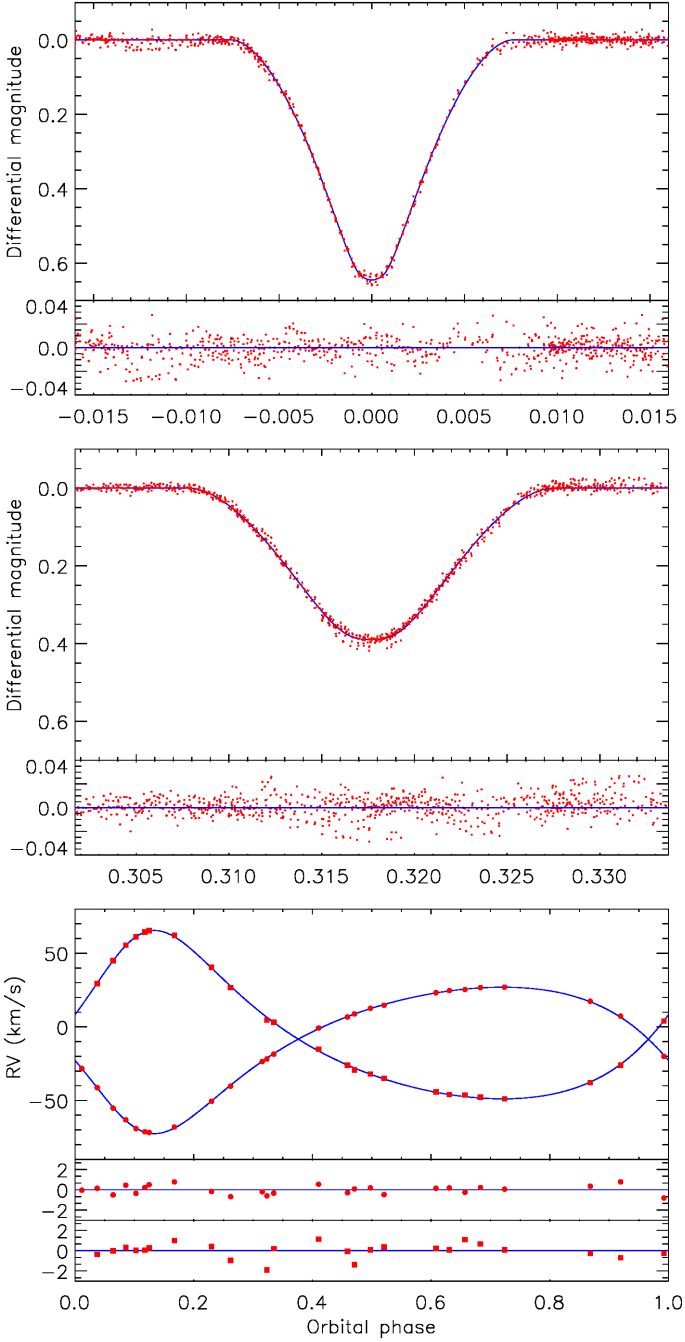}
\caption{\label{fig:lcrv} SuperWASP light curve of the primary (top) and 
secondary eclipse (middle) and \citet{Griffin13obs} RVs (bottom) of LL\,Aqr.
The datapoints are shown in red and the solid lines show the best fits found 
using {\sc jktebop}. The residuals are shown below each fit.} \end{figure}

\begin{figure} \includegraphics[width=\columnwidth,angle=0]{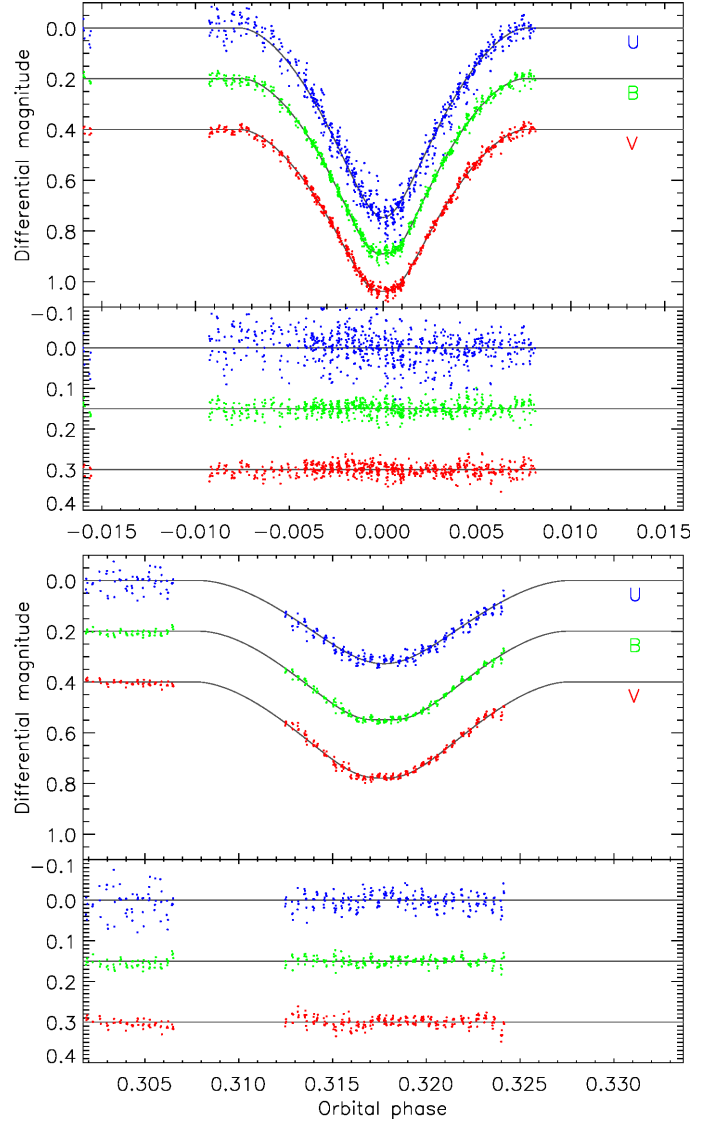}
\caption{\label{fig:iblc} $UBV$ light curves of the primary (top) and 
secondary eclipse (bottom) of LL\,Aqr. The datapoints are shown in blue 
($U$), green ($B$) and red ($V$), and the solid lines show the best fits found 
using {\sc jktebop}. The residuals are shown below each fit.} \end{figure}

The component stars of LL\,Aqr are well-detached and almost spherical, making the system well-suited to analysis using the {\sc jktebop} code\footnote{{\sc jktebop} is written in {\sc fortran77} and the source code is available at {\tt http://www.astro.keele.ac.uk/jkt/codes/jktebop.html}}, as originally presented by \citet{Me++04mn2} and with significant extensions by \citet{Me++07aa} and \citet{Me08mn}. {\sc jktebop} is based on the {\sc ebop} code \citep{Etzel81conf,PopperEtzel81aj} with the {\sc nde} model \citep{NelsonDavis72apj}. It represents the star and planet as biaxial spheroids but adopts the spherical approximation for the calculation of light lost during eclipse.

Important parameters of the {\sc jktebop} model are the fractional radii of the two stars, 
\begin{equation}
r_{\rm A} = \frac{R_{\rm A}}{a} \qquad \qquad r_{\rm B} = \frac{R_{\rm B}}{a}
\end{equation}
where $a$ is the orbital semimajor axis, and $R_{\rm A}$ and $R_{\rm B}$ are the true radii of the stars. We fitted for their sum and ratio:
\begin{equation}
r_{\rm A} + r_{\rm B} \qquad \qquad k = \frac{r_{\rm B}}{r_{\rm A}} = \frac{R_{\rm B}}{R_{\rm A}}
\end{equation}
plus the orbital inclination ($i$), the central surface brightness ratio of the two stars ($J$), and the nuisance parameter the out-of-eclipse magnitude. 

The orbital period (\Porb) and the time of primary mid-eclipse ($T_{\rm pri}$) were also included as fitted quantities. The orbital ephemeris was further constrained by including the six published times of minimum light as observational quantities, using the method of \citet{Me++07aa}.

For limb darkening we adopted the quadratic law with coefficients\footnote{\reff{The adopted limb darkening coefficients were the average of the values for the Gunn $g$ and $r$ passbands.}} taken from \citet{ClaretBloemen11aa}. The precise values of the limb darkening coefficients are not important to the results, but the light curve contains sufficient information for the linear coefficient for each star ($u_{\rm A}$ and $u_{\rm B}$) to be fitted. The quadratic coefficients ($v_{\rm A}$ and $v_{\rm B}$) were held fixed.

We checked for the possibility of contaminating ``third'' light, motivated by the detection of a third signal in two of the 12 spectra obtained by I08, by including it as a fitted parameter. We obtained a value which was very close to zero and much smaller than its uncertainty, so we fixed third light at zero for subsequent calculations.

LL\,Aqr presents a significant orbital eccentricity which we accounted for by fitting for the quantities $e\cos\omega$ and $e\sin\omega$, where $e$ is eccentricity and $\omega$ is the longitude of periastron. This is because $e$ and $\omega$ can be strongly correlated \citep[e.g.][]{Pavlovski+09mn}, compromising their utility as fitted parameters. Light curves of eccentric dEBs typically allow the quantity $e\cos\omega$ to be determined precisely, as it primarily affects the time difference between primary and secondary eclipse \citep[e.g.][]{Kopal59book}. The main indicator of $e\sin\omega$ is the ratio of the eclipse durations, which are generally less precise. The inclusion of RVs in an analysis can allow a significantly improved measurement of $e\sin\omega$ \citep[see][]{Wilson79apj}. We therefore decided to modify {\sc jktebop} to allow simultaneous fitting to the SuperWASP light curve and the RVs of both stars\footnote{The modified {\sc jktebop} code was also used by \citet{Frandsen+13aa} in the study of the giant dEB system KIC\,8410637.}. 

We included the velocity amplitudes of the two stars ($K_{\rm A}$ and $K_{\rm B}$) and the systemic velocities ($\gamma_{\rm A}$ and $\gamma_{\rm B}$) as fitted parameters. \reff{We allowed the stars to have different systemic velocities in order to avoid possible systematic effects due to mismatch between their spectra and the template used to observe the cross-correlation function \citep[see also][]{PopperHill91aj}. The fitted values of $\gamma_{\rm A}$ and $\gamma_{\rm B}$ differ by an acceptable 1.5$\sigma$. We checked the consequences of this approach by computing an alternative solution where the two stars have a common systemic velocity. This yielded values for $K_{\rm A}$, $K_{\rm B}$ and $e\sin\omega$ which were larger by 0.3$\sigma$, and had a smaller effect on all other parameters.}

Measurement errors are not available for the RVs (they were all accorded unit weight by \citealt{Griffin13obs}), and are far too small for the SuperWASP data, so all datapoints in individual datasets were assigned the same weight. We incorporated the iterative adjustment of weights for the individual datasets in {\sc jktebop}, in order to appropriately combine the different sources of information. This was done by forcing the reduced $\chi^2$ (\chir) for each dataset to be unity; by doing this we are implicitly assuming that the best fit we find is a good representation of the data.

Uncertainties were calculated using a Monte Carlo (MC) approach \citep{Me++04mn2} with 1000 synthetic datasets per run. Whilst this does a good job for well-behaved data \citep{Me+05mn}, we have found that it underestimates the true uncertainties in the presence of correlated noise such as occurs in the SuperWASP data \citep{Me+11mn2}. We therefore ran residual-permutation (RP) simulations \citep[see][]{Me08mn}, after modifying the implementation of this algorithm in {\sc jktebop} such that the residuals are permuted only within datasets. To account for the much smaller number of RV compared to photometric datapoints, we set the residuals for smaller datasets to cycle multiple times whilst the residuals for the largest dataset cycle exactly once.

The final results of this analysis are contained in Table\,\ref{tab:lcsw} and are precisely and accurately measured. The uncertainties on individual parameters were assessed in three ways: from the covariance matrix, from MC and from RP. The parameters primarily determined by the photometry ($r_{\rm A}$, $r_{\rm B}$, $J$, $i$, \Porb, $T_{\rm pri}$) show uncertainties from the RP analysis which are roughly twice those from the MC approach. The formal errors agree well with the MC results, which is sensible because the solution does not suffer from the strong correlations between parameters which formal errors do not account for. 

We find an orbital ephemeris of 
$$ T_{\rm pri} {\rm (HJD/UTC)} = 2455098.54955 (12) + 20.178321 (6) \times E $$
where each bracketed quantity represents the uncertainty in the last digit of the preceding number, and $E$ is the number of orbital cycles since the reference epoch. This orbital period is in good agreement with the value obtained by \citet{Griffin13obs}. The period found by I08 has a higher quoted precision than our own, despite being based on fewer data, and is in formal but unexceptional disagreement with our result.

\begin{table} \centering \caption{\label{tab:lcsw} Individual parameters of the fits 
to the \reff{SuperWASP light curve} of LL\,Aqr, including several alternative sets of errorbars.}
\begin{tabular}{l c c c c} \hline \hline
Parameter               & Value & Formal error & MC error & RP error \\
\hline
$r_{\rm A}+r_{\rm B}$   & 0.05692 & 0.00016 & 0.00016 & 0.00034 \\
$k$                     & 0.7535  & 0.0038  & 0.0038  & 0.0063  \\
$i$ ($^\circ$)          & 89.550  & 0.021   & 0.022   & 0.033   \\
$J$                     & 0.767   & 0.020   & 0.020   & 0.036   \\
$u_{\rm A}$             & 0.463   & 0.046   & 0.048   & 0.068   \\
$u_{\rm B}$             & 0.503   & 0.057   & 0.056   & 0.108   \\
$e\cos\omega$           &$-$0.28803&0.00008 & 0.00009 & 0.00012 \\
$e\sin\omega$           & 0.1313  & 0.0021  & 0.0023  & 0.0021  \\
$K_{\rm A}$ (\kms)      & 49.72   & 0.12    & 0.12    & 0.09    \\
$K_{\rm B}$ (\kms)      & 57.19   & 0.20    & 0.19    & 0.20    \\
$\gamma_{\rm A}$ (\kms) & $-$8.45 & 0.09    & 0.10    & 0.03    \\
$\gamma_{\rm B}$ (\kms) & $-$8.18 & 0.15    & 0.16    & 0.04    \\
\hline
$v_{\rm A}$             & 0.39    &         &         &         \\
$v_{\rm B}$             & 0.29    &         &         &         \\
$r_{\rm A}$             & 0.03246 &         & 0.00013 & 0.0025  \\
$r_{\rm B}$             & 0.02446 &         & 0.00008 & 0.0016  \\
Light ratio             & 0.4317  &         & 0.0019  & 0.0043  \\
$e$                     & 0.31654 &         & 0.00086 & 0.00080 \\
$\omega$ (degrees)      & 155.50  &         & 0.38    & 0.36    \\
rms (mmag)              & 9.4     &         &         &         \\
\hline \end{tabular} \tablefoot{\reff{The upper part of the table 
gives the fitted parameters and the lower part of the table gives 
fixed or calculated quantities.}} \end{table}

\begin{table*} \centering \caption{\label{tab:lcib} Individual parameters of the fits 
to the $UBV$ light curves of LL\,Aqr, including two alternative sets of errorbars.}
\begin{tabular}{l c c c c c c c c c} \hline \hline
\ & \multicolumn{3}{c}{$U$ light curve} & \multicolumn{3}{c}{$B$ light curve}  & \multicolumn{3}{c}{$V$ light curve} \\
Parameter               & Value & MC error & RP error & Value & MC error & RP error & Value & MC error & RP error \\
\hline
$r_{\rm A}+r_{\rm B}$   & 0.05792 & 0.00030 & 0.00064 & 0.05640 & 0.00014 & 0.00035 & 0.05722 & 0.00014 & 0.00042 \\
$k$                     & 0.870   & 0.048   & 0.083   & 0.7570  & 0.0018  & 0.0034  & 0.7564  & 0.0020  & 0.0041  \\
$i$ ($^\circ$)          & 89.273  & 0.071   & 0.115   & 89.617  & 0.017   & 0.031   & 89.535  & 0.015   & 0.029   \\
$J$                     & 0.6036  & 0.0054  & 0.0129  & 0.6781  & 0.0021  & 0.0059  & 0.7395  & 0.0020  & 0.0095  \\
$e\cos\omega$           &$-$0.28790&0.00009 & 0.00022 &$-$0.28796&0.00004 & 0.00009 &$-$0.28806&0.00003 & 0.00009 \\
\hline
$e\sin\omega$           & 0.1313  &         &         & 0.1313  &         &         & 0.1313  &         &         \\
$u_{\rm A}$             & \reff{0.42} &     &         & \reff{0.46} &     &         & \reff{0.33} &     &         \\
$u_{\rm B}$             & \reff{0.34} &     &         & \reff{0.29} &     &         & \reff{0.32} &     &         \\
$v_{\rm A}$             & \reff{0.59} &     &         & \reff{0.54} &     &         & \reff{0.38} &     &         \\
$v_{\rm B}$             & \reff{0.24} &     &         & \reff{0.24} &     &         & \reff{0.30} &     &         \\
$r_{\rm A}$             & 0.03098 & 0.00089 & 0.00148 & 0.03210 & 0.00007 & 0.00015 & 0.03258 & 0.00007 & 0.00020 \\
$r_{\rm B}$             & 0.02694 & 0.00073 & 0.00126 & 0.02430 & 0.00008 & 0.00019 & 0.02464 & 0.00008 & 0.00022 \\
Light ratio             & 0.439   & 0.051   & 0.089   & 0.3788  & 0.0024  & 0.0059  & 0.4189  & 0.0028  & 0.0065  \\
$e$                     & 0.31643 & 0.00008 & 0.00020 & 0.31648 & 0.00004 & 0.00008 & 0.31657 & 0.00003 & 0.00008 \\
$\omega$ (degrees)      & 155.484 & 0.007   & 0.016   & 155.489 & 0.003   & 0.007   & 155.496 & 0.002   & 0.007   \\
Light curve rms (mmag)  & 33      &         &         & 15      &         &         & 13      &         &         \\
\hline \end{tabular} 
\tablefoot{\reff{The upper part of the table gives the fitted parameters 
and the lower part of the table gives fixed or calculated quantities.}}
\end{table*}

We also modelled the $UBV$ light curves presented by I08, each of which contains 1925 datapoints with full coverage of the primary eclipse but only partial coverage of the secondary eclipse. We fixed the orbital ephemeris at that determined to high precision from the preceding analysis, and also fixed $e\sin\omega$ at the value found above to make up for not including RVs in this step. The limb darkening coefficients were also not adjusted after it was established that the light curves have insufficient precision to constrain them to within useful limits. The best fits are shown in Fig.\,\ref{fig:iblc}. Errorbars were calculated via the MC and RP algorithms and both are shown in Table\,\ref{tab:lcib}. The errorbars for $e$ and $\omega$ are underestimated because $e\sin\omega$ was fixed in these analyses. For all three light curves the RP uncertainties are at least a factor of two larger than the MC uncertainties, showing that dedicated observations can be as strongly affected by correlated noise as robotic survey data. \reff{We adopted the larger of the MC and RP errorbars for each parameter.}



\begin{table} \centering \caption{\label{tab:lcfinal} 
Final photometric and spectroscopic parameters for LL\,Aqr.}
\begin{tabular}{l r@{\,$\pm$\,}l} \hline \hline
Parameter               & \mc{Value} \\
\hline
$r_{\rm A}$             & 0.03226   & 0.00015 \\
$r_{\rm B}$             & 0.02448   & 0.00019 \\
$i$ ($^\circ$)          & 89.560    & 0.033   \\
$e$                     & 0.31654   & 0.00086 \\
$\omega$ (degrees)      & 155.50    & 0.38    \\
$K_{\rm A}$ (\kms)      & 49.72     & 0.12    \\
$K_{\rm B}$ (\kms)      & 57.19     & 0.20    \\
Light ratio ($U$)       & 0.327     & 0.089   \\
Light ratio ($B$)       & 0.379     & 0.006   \\
Light ratio ($V$)       & 0.419     & 0.007   \\
\hline \end{tabular} \end{table}

For calculating the final photometric and spectroscopic parameters we adopted the weighted mean of the individual values of $r_{\rm A}+r_{\rm B}$, $k$, $i$, $r_{\rm A}$ and $r_{\rm B}$. The results for the $U$ light curve are outwardly slightly discrepant, but in fact their large uncertainties make them consistent with the other values to within 1-2$\sigma$. We calculated weighted means both with and without the $U$ results and found them to differ insignificantly, so we adopt those including the $U$ results. The \chir\ of the values around the weighted means are 3.5 for $i$, 1.5 for $r_{\rm A}$ and 1.7 for $r_{\rm B}$. These disagreements are not big enough to indicate a problem with the data (or analysis), and have been accounted for by inflating the uncertainties in the affected parameters to enforce $\chir = 1$. The final parameters are given in Table\,\ref{tab:lcfinal}; those not already discussed in this paragraph have been adopted unchanged from individual analyses. The final light ratios in $UBV$ were calculated by repeating the fits to these light curves whilst fixing the geometrical parameters to the values given in Table\,\ref{tab:lcfinal}. \refff{A model light curve calculated for the geometrical properties in Table\,\ref{tab:lcfinal} shows that the primary eclipse is total but that the secondary eclipse is partial.}



\section{Physical properties and distance}                                                                                         \label{sec:absdim}

The physical properties of LL\,Aqr have been calculated from the values of \Porb, $r_{\rm A}$, $r_{\rm B}$, $i$, $e$, $K_{\rm A}$ and $K_{\rm B}$ measured in the preceding section. This was done using the {\sc jktabsdim} code \reff{originally developed by \citet{Me++05aa}}, which propagates the uncertainty in each input parameter by a perturbation approach. Results are calculated for each input parameter, and plus and minus its errorbar, yielding the effect of its uncertainty on each output parameter. The individual contributions are then added in quadrature to generate the final errorbar for each output quantity. \reff{The adopted physical constants are listed in \citet{Me11mn}.}

The masses and radii of the two stars (Table\,\ref{tab:absdim}) are all measured to high precision (less than 1\%), thanks to the high quality of the RVs from \citet{Griffin13obs} and the large quantity of photometric observations\footnote{The properties of LL\,Aqr have been entered into the DEBCat catalogue of well-studied detached eclipsing binary star systems: {\tt http://www.astro.keele.ac.uk/jkt/debcat/}}. LL\,Aqr is therefore an excellent check and calibrator of theoretical models in the region of the Sun; the mass and radius of the secondary component are almost identical to those of the Sun. This will be exploited in the next section. The synchronous rotational velocities of the stars are only 3\kms. Tidal effects move the rotational velocities towards a pseudosynchronous state, i.e.\ synchronous with the orbital motion at periastron when the stars have their smallest separation \citep{Hut81aa}. The pseudosynchronous rotational velocities of the stars are $6.64 \pm 0.04$\kms\ for the primary star and $5.04 \pm 0.04$\kms\ for the secondary star. 

The distance to the system is also straightforwardly measurable. For this we adopt the \Teff\ values from I08 and the 2MASS $JHK$ apparent magnitudes\footnote{\refff{The 2MASS magnitudes are $J = 8.145 \pm 0.023$, $H = 7.872 \pm 0.033$ and $K_s = 7.819 \pm 0.023$. These observations were taken at epoch JD 2451315.9260, corresponding to phase 0.54 for LL\,Aqr, so are representative of its out-of-eclipse brightness.}} \citep{Skrutskie+06aj}. I08 measured standard magnitudes for LL\,Aqr of $V = 9.206$, $B-V = 0.559$ and $U-B = 0.085$. The $V$-band measurement does not agree well with the values of $V = 9.32 \pm 0.02$ and $9.86 \pm 0.03$ obtained from {\it Tycho} observations \citep{Hog+97aa}. There is a multitude of different (but often related) alternatives listed by {\it Vizier}\footnote{{\tt http://vizier.u-strasbg.fr/viz-bin/VizieR}} and these show a spread from $V = 9.2$ to $V = 9.4$ so cannot break the deadlock. The $B$ magnitudes are (surprisingly) more concordant.

Our favoured method for determining the distance to the system is via surface brightness relations \citep{Me++05aa}, using the empirical calibrations provided by \citet{Kervella+04aa}. When adopting this method, the $UBV$ magnitudes from I08 and the 2MASS $JHK$ magnitudes, we were able to find a consistent distance of 135--139\,pc in each of the $BVJHK$ passbands. The $U$ passband yields a distance about 6\,pc shorter, which is well within its uncertainty. We were able to obtain distances to the individual stars in $UBV$ as we possess flux ratios between the stars in these passbands; their consistency shows that the \Teff\ measurements (or at least their ratio) from I08 are reliable.

For our distance measurement we required an interstellar reddening excess of $E(B-V) = 0.12 \pm 0.05$ (conservative errorbar) to align the distances found in the $BV$ and $JHK$ passbands. We adopt the final distance value of $137.8 \pm 2.7$\,pc from the $K$ band, as this value is least affected by the uncertainties in \Teff\ and $E(B-V)$. \reff{The revised {\it Hipparcos} parallax \citep{Vanleeuwen07aa} of $8.62 \pm 1.15$\,mas corresponds to a distance of \er{116}{18}{14}\,pc, which agrees with our final distance measurement to within 1.4$\sigma$.}

\begin{table} \label{tab:absdim}
\caption{The physical properties of the LL\,Aqr system.}
\centering
\begin{tabular}{l r@{\,$\pm$\,}l r@{\,$\pm$\,}l} \hline \hline
Parameter                     &    \mc{Star A}   &    \mc{Star B}    \\ \hline
Orbital separation (\Rsun)    &\multicolumn{4}{c}{$40.46 \pm 0.009$} \\
Mass (\Msun)                  &  1.167  & 0.009  &  1.014  & 0.006   \\
Radius (\Rsun)                &  1.305  & 0.007  &  0.990  & 0.008   \\
$\log g$ [cm\,s$^{-2}$]       &  4.274  & 0.004  &  4.453  & 0.007   \\
\Vsync\ (\kms)                &  3.27   & 0.02   &  2.48   & 0.02    \\
\Vsini\ (\kms)                &  4      & 2      &  4      & 2       \\[3pt]
\Teff (K)                     &  6680   & 160    &  6200   & 160     \\
$\log(L/\Lsun)$ $^1$          &  0.483  & 0.042  &  0.114  & 0.045   \\
\Mbol\ $^1$                   &  3.54   & 0.10   &  4.47   & 0.11    \\
Distance (pc)                 &  \multicolumn{4}{c}{$137.8 \pm 2.7$} \\
\hline \end{tabular}
\tablefoot{$^1$ Calculated assuming $\Lsun = 3.844${$\times$}10$^{26}$\,W
\citep{Bahcall++95rvmp} and $\Mbol\sun = 4.75$ \citep{Zombeck90book}.}
\end{table}


\section{Comparison with theoretical stellar models}                                                                               \label{sec:models}

\begin{figure} \includegraphics[width=\columnwidth,angle=0]{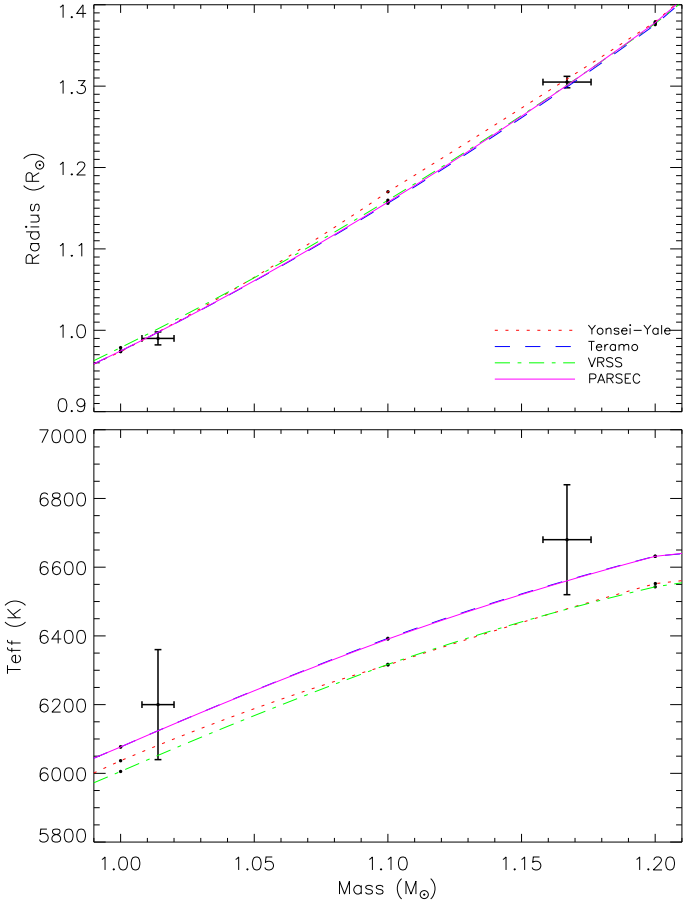}
\caption{\label{fig:model} Mass--radius and mass--\Teff\ plots comparing the 
physical properties of LL\,Aqr (points with errorbars) with the predictions 
of four sets of theoretical stellar models (points without errorbars show 
the tabulated predictions and lines show a quadratic interpolation between
these for each set of models).} \end{figure}


\reff{We have compared the physical properties of LL\,Aqr} to several sets of tabulated theoretical predictions in the mass--radius and mass--\Teff\ planes. Such comparisons are informative because the physical properties of the two stars are precisely measured and very different: the unevolved secondary star is a good indicator of bulk metal abundance, $Z$, whereas the properties of the slightly evolved primary indicate the age of the system, $\tau$, for a given $Z$. The long orbital period also means that tidal interactions have had an insignificant effect on the evolution of the two stars, so they are reliably representative of single stars.

\refff{One important caveat is that the measured \Teff\ values of the two stars are not independent. I08 determined the \Teff\ of the primary star from its colour index, and fixed this when analysing the light curve. The \Teff\ of the secondary star was thus determined relative to that of the primary. In the case of LL\,Aqr the ratio of the \Teff s is well-determined, but the absolute values are less precise. This situation is common in the analysis of dEBs \citep[e.g.][]{Claret03aa,MeClausen07aa}.}

Armed with the Yonsei-Yale models \citep{Demarque+04apjs} and adopting a scaled-solar chemical composition, we find that predictions for $Z = 0.01$ and $\tau = 2.9$\,Gyr match the masses and radii well but under-predict the \Teff\ values by 1$\sigma$. Adopting a lower $Z = 0.007$ and $\tau = 2.6$\,Gyr matches the mass and radius of the primary star and the \Teff\ values of both stars, but predicts a modestly larger radius for the mass of the secondary star. \refff{The mean of the two sets of predictions provides a good match to the properties of LL\,Aqr.}

The equivalent VRSS \citep{Vandenberg++06apjs} models for $Z = 0.01$ and $\tau = 2.5$\,Gyr also match both masses and radii well, but under-predict the temperatures by 1$\sigma$. The Teramo \citep{Pietrinferni+04apj} models for $Z = 0.01$ and the PARSEC models \citep{Bressan+12mn} for $Z = 0.008$, both for age $\tau = 2.5$\,Gyr, are almost identical and good matches to both stars. They fit the radii well and under-predict the \Teff\ values by about 0.7$\sigma$. \reff{They are therefore in formally good agreement with the properties of LL\,Aqr.}

\begin{figure*} \includegraphics[width=\textwidth,angle=0]{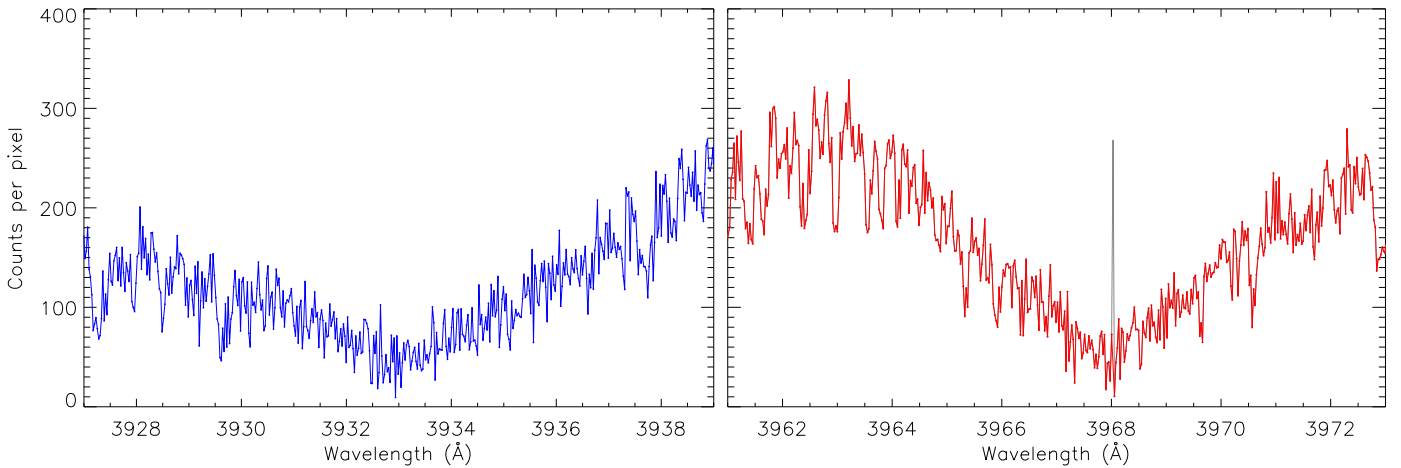}
\caption{\label{fig:cahk} The spectrum of LL\,Aqr in the region of the Ca
H and K lines, showing the lack of chromospheric emission in the line cores.
One pixel suffered from a cosmic ray event and has been greyed out.} \end{figure*}

\refff{We conclude that it is possible to find theoretical predictions which are consistent with the physical properties of the LL\,Aqr system for metallicities of $Z = 0.007$--$0.010$ and ages in the interval 2.5--2.9\,Gyr, although the majority of theoretical models tested would prefer the stars to be cooler by about 100\,K.} The dereddened $B-V$ indices found in Section\,\ref{sec:anal} yield \Teff\ values \citep{Boyajian+13apj} which are lower than those found by I08, but not by a statistically significant amount. \refff{A definitive confrontation with theoretical model predictions could be performed once precise spectroscopic \Teff s and chemical abundances have been obtained for the two stars.}


\section{Indications of stellar activity}                                                                                        \label{sec:activity}

\begin{table} \centering \caption{\label{tab:specdata} 
\reff{\'Echelle spectrum of LL\,Aqr. The data  
are available in their entirety at the CDS.}}
\begin{tabular}{c c c} 
\hline \hline
Order & Wavelength (\AA) & Counts \\ 
\hline
1 & 3830.4316 & 53.3 \\
1 & 3830.4579 & 60.1 \\
1 & 3830.4842 & 47.8 \\
1 & 3830.5106 & 38.2 \\
1 & 3830.5369 & 73.5 \\
\hline \end{tabular} \end{table}

Low-mass stars often exhibit signs of stellar activity such as modulated brightness due to starspots, enhanced UV and X-ray flux, and emission lines at optical wavelengths. We checked for each of these possibilities.

Firstly, the residuals of the SuperWASP data versus the best fit were subjected to a period analysis using the {\sc period04} package \citep{LenzBreger04iaus}. A Fourier transform for periods below 1\,d shows no evidence for sinusoidal variation, with a 3$\sigma$ limit of 3\,mmag. The limits for the individual stars are 5\,mmag (primary) and 7\,mmag (secondary) as they both contribute to the flux in the SuperWASP passband.

Secondly, LL\,Aqr was not detected in the ROSAT Point Source Catalogue \citep{Voges+99conf} to a limit of 0.1\,counts\,s$^{-1}$ in the 0.1--2.5\,keV band. It has been detected in the FUV and NUV bands by the GALEX satellite \citep{Morrissey+07apjs}, at magnitudes 20.45 and 14.18 respectively. These measurements do not point to an enhanced X-ray or UV flux arising from the system.

Finally, the best indicators of chromospheric activity at optical wavelengths are the calcium H and K lines at 3933\,\AA\ and 3967\,\AA\ \citep{Wilson68apj,Duncan+91apjs}. These were not in the wavelength range of the spectra obtained by I08, so we acquired a single spectrum of LL\,Aqr using the 3\,m Shane Telescope at the Lick Observatory, US. We used the Hamilton \'echelle spectrograph to obtain a spectrum covering the full optical range at a resolving power of approximately 60\,000 \reff{(Table\,\ref{tab:specdata})}. The Ca H and K lines show no emission, except for a mischievously placed cosmic ray event (Fig.\,\ref{fig:cahk}). Whilst Ca H and K emission is generally weak at the temperatures of the components of LL\,Aqr \citep[e.g.][]{Knutson++10apj}, spot and chromospheric activity is seen in our own Sun for which the mass and radius of LL\,Aqr\,B is a good match. \reff{We also see no emission at H$\alpha$.} We conclude that LL\,Aqr shows no evidence of stellar activity or magnetism.

We also used the spectrum to measure the projected rotational velocities of the two stars. The spectrum was taken at phase 0.011 (just after the end of primary eclipse) when the velocity difference of the stars was 43\kms\ and their spectral lines were well separated. We first measured an instrumental broadening of 0.09\,\AA\ from the emission lines in the thorium-argon calibration exposures. We then modelled selected spectral lines using the {\sc uclsyn} code \citep{Smalley++01,Smith92phd}, finding $\Vsini = 4 \pm 2$\kms\ for the primary star and $\Vsini = 4 \pm 2$\kms\ for the secondary star. These values are consistent with both synchronous and pseudo-synchronous rotation.


\section{Conclusions}                                                                                                                \label{sec:conc}

The eclipsing nature of the binary system LL\,Aqr was discovered from {\it Hipparcos} photometry. Its joint properties of a solar-twin secondary star and a long orbital period make it ideal for studying the evolution of solar-type stars where the complicating effects of tidal interactions are weak.

We measured the physical properties of LL\,Aqr to high precision based on high-quality published radial velocity measurements from \citet{Griffin13obs} and extensive photometry obtained by the SuperWASP instruments located at La Palma and South Africa. The {\sc jktebop} code was modified to simultaneously fit RVs and light curves for this work. These data were augmented by published $UBV$ light curves from I08, allowing the SuperWASP results to be checked and refined. The resulting masses and radii are model-independent and have precisions of better than 1\%.

Theoretical models are able to match the measured physical properties of LL\,Aqr for an approximately half-solar chemical composition and an age of 2.5\,Gyr. More sensitive tests of stellar theory could be achieved through obtaining precise spectroscopic \Teff\ and chemical abundance measurements. LL\,Aqr displays no evidence of spot activity or chromospheric emission, and both components are rotating slowly.

\reff{Tidal effects act to alter the rotational velocities and orbital eccentricity of the stars. Using the tidal friction theory of \citet{Zahn77aa} we find a rotational synchronisation timescale of approximately 1.5\,Gyr and an orbital circularisation timescale which is greater by three orders of magnitude. Both timescales are consistent with the properties of LL\,Aqr at an age of 2.5\,Gyr: the stars have rotational velocities consistent with both synchronous and pseudo-synchronous rotation, and the orbit has not been circularised.}

The comparison with theoretical models shows that the components of LL\,Aqr do not exhibit the anomalously large radii typical for low-mass dEBs \citep{Ribas06apss,Ribas+08conf,Lopez07apj,Me09mn}. Such inflated radii are explained by invoking fast rotation in short-period binary systems due to tidal effects. This enhances the magnetic activity of the star, which in turn inhibits convective heat transport and thus causes the star to be larger and cooler. A prediction of this concept is that long-period dEBs will not show inflated radii. LL\,Aqr does not, in line with KIC\,6131659 ($\Porb = 17.5$\,d, \citealt{Bass+12apj}) and RW\,Lac ($\Porb = 10.4$\,d, \citealt{Lacy+05aj}). LSPM\,J1112+7626 ($\Porb = 41.0$\,d, \citealt{Irwin+11apj}) is an M-dwarf dEB which shows spot activity and inflated radii; its discord with the picture outlined above could easily be due to a young age.


\begin{acknowledgements}
JS acknowledges funding from STFC in the form of an Advanced Fellowship. We are grateful to the SuperWASP Consortium for obtaining the main light curve used in this work, to Roger Griffin and Serdar Evren for supplying published observations of LL\,Aqr, to Kelsey Clubb for obtaining the Hamilton spectrum of LL\,Aqr, and to Barry Smalley and Roger Griffin for discussions. We thank the anonymous referee for a helpful report. The following internet-based resources were used in research for this paper: the NASA Astrophysics Data System; and the SIMBAD database operated at CDS, Strasbourg, France.
\end{acknowledgements}


\bibliographystyle{aa}

\end{document}